\documentclass[11pt]{article}
\usepackage{geometry}                % See geometry.pdf to learn the layout options. There are lots.
\usepackage{graphicx}
\usepackage{amsmath, amsthm, amsfonts}
\usepackage{amssymb}

\newcommand{\sect}[1]{\setcounter{equation}{0}\section{#1}}

\usepackage{color}

\usepackage{epstopdf}
\title{Classical ladder functions for Rosen-Morse and curved Kepler-Coulomb systems
}
\author{
L.  Delisle-Doray$^a$, V. Hussin$^b$, \c{S}. Kuru$^c$, J. Negro$^d$ 
\bigskip
\\
\noindent
$^a$D\'epartement de Physique \& Centre de Recherches Math\'ematiques,
\\
Universit\'e de Montr\'eal, QC, H3C 3J7, Canada
\\
\noindent
$^b$D\'epartement de Math\'ematiques et de Statistique \& Centre de Recherches Math\'ematiques,
\\
Universit\'e de Montr\'eal, QC, H3C 3J7, Canada
\\
\noindent
$^c$Department of Physics, Faculty of Science, Ankara
University, 06100 Ankara, Turkey
\\ 
 \noindent
$^d$Departamento de F\'{\i}sica Te\'orica, At\'omica y
\'Optica, Universidad de Valladolid,\\  47011 Valladolid, Spain
}
\begin{document}
\maketitle

\begin{abstract}

Ladder functions in classical mechanics are defined in a similar way as
ladder operators in the context of quantum mechanics. 
In the present paper, we develop a new
method for obtaining ladder functions of one dimensional systems by means of 
a product of two `factor functions'. We apply 
this method to the curved Kepler-Coulomb and Rosen-Morse II systems whose ladder functions were not found yet. The ladder functions here obtained
are applied to get the motion of the system.

\end{abstract}

\bigskip\bigskip 

\noindent
PACS:\quad

 \medskip


In quantum mechanics the knowledge of symmetry
operators $\hat S$ that commute with the Hamiltonian operator $\hat H$
of the system is very important. The set of symmetry operators closes, under the composition operation,
a `symmetry algebra'. This type of operators gives useful information
on the system, in particular they explain the degeneracy of the energy
levels and also they can supply the spectrum. A second type of interesting operators are called ladder operators $\hat A$.
They do not commute with the Hamiltonian, but satisfy simple commutation rules
of conformal type: $[\hat H, \hat A] = \alpha(\hat H) \hat A$, where $\alpha(\hat H)$ designs
a certain function of $\hat H$. When such operators exist, they connect eigenspaces
of different energies; the paradigmatic case is given by the lowering and raising operators of the harmonic oscillator $a^\pm$. If we include both, symmetry and ladder operators, we
get a `spectrum generating algebra' \cite{Barut65,Dothan65,Dothan70}.

In the frame of the Hamiltonian formalism of classical mechanics, the symmetries
are generated by functions $S(x,p)$ of the canonical variables such that
the Poisson bracket with the Hamiltonian function $H(x,p)$ vanishes:
$\{H,S\}=0$. The surfaces defined by the values of such functions, $S(x,p)= s_0$,
are made of trajectories of the system and any trajectory must be included
in one of such surfaces. Although much less known,  one  can also define
ladder functions $A(x,p)$ for classical systems as those satisfying the corresponding
Poisson bracket with the Hamiltonian: $\{H,A\} = i \, \alpha(H) A$. In the classical
context, the ladder functions supply information on the motion of the
system.

The aim of this paper is to find the ladder functions of two families of one
dimensional systems known as Rosen-Morse II (RMII) \cite{cooper00,RM} and curved Kepler-Coulomb (KC) \cite{Stevenson41,Infeld45,Ballesteros09,ranada99}. 
Ladder functions for other one dimensional systems have been computed in previous
works \cite{kuru08,kuru12}, but they were still missing for the RMII and curved KC systems, in particular. 
The ladder functions of these two systems are considerably more elaborate, 
and the purpose of this work is to introduce a procedure to get them.
In this way, we want to complete the knowledge of this type of functions for 
the whole
list of classical systems corresponding to the quantum systems
solved by means of the factorization method as given by Infeld-Hull \cite{cooper00,infeld51}. 
  
The structure of the paper is as follows. In Section 2, we introduce the ladder
functions and some of their basic properties. Next, in Section 3, we show
our method to compute ladder functions as a product of two `factor functions'.
Then, in Section 4 we apply this method to the RMII and show that it includes the well known results for the
P\"oschl-Teller system as a particular case. Next, in Section 5 we deal with the
curved KC system.
Section 6 considers the limit from curved to flat KC systems and the
case with zero angular momentum in order to get the formulas
obtained previously by other methods. Some remarks will end the paper in
the last section.

%
%
%We give a method for obtaining a ladder function of classical mechanics. It consists in combining different well-chosen functions of the phase space to obtain a proper ladder function for a given potential. We then apply the method to the case of the Rosen-Morse and Kepler-Coulomb potentials, for which no such function has been found yet. To do so we will use three tools (representation formula, signature, contribution), which we will introduce and explain in the next sections.

\sect{Basic theory}

\subsection{Definition of ladder functions}

Let $H$ be a one-dimensional Hamiltonian with canonical coordinates $(x,p)$ defined as
\begin{equation}
H = p^2 + V(x)\,.
\end{equation}
We want to find two complex functions $A^\pm(x,p)$ defined on (some part of) the phase space $(x,p)$,  that together with the Hamiltonian $H(x,p)$ satisfy the following Poisson bracket  algebra \cite{kuru08,kuru12,campoamor12}: 
\begin{equation}
    \{H,A^{\pm}\} = \mp i \alpha(H)A^{\pm}, \label{GHA1}
\end{equation}
\begin{equation}
    \{A^+,A^-\} = i\beta(H),  \label{GHA2}
\end{equation}
where, in principle, $\alpha(H)$ and $\beta(H)$ are real functions which depend on $H$. We also may assume that $\alpha(H)$ be positive. Such an algebra is a classical analog of the generalized Heisenberg algebra (GHA), which is satisfied by ladder operators, with respect to commutators, in quantum mechanics \cite{marquette11,bagarello18}. Since (\ref{GHA2}) can be deduced from (\ref{GHA1}), and since the ladder functions $A^\pm$ can in general be taken to be complex conjugate of each other, $A^+=(A^-)^*$ (at least for bound states), only one of the equations (\ref{GHA1}) will be relevant.
%\begin{equation}
%    \{H,A^{-}\} = i \alpha(H)A^{-}, \label{GHA relevant}
%\end{equation}
%so that whenever we refer to only one ladder function, it is understood that we mean $A^-$. 
From (\ref{GHA1}) we can also deduce a type of factorization in the following form:

\begin{equation}
    \delta(H) = A^{+}A^{-}, \label{Factorization}
\end{equation}
where $\delta(H)$ is a certain function depending only on $H$.
This is the reason why the ladder functions are also referred as the classical counterpart to the factorization method in quantum mechanics.
We will note however that (\ref{Factorization}) is actually weaker than its quantum analog  because of the commutativity of the product of phase-space functions. We will refer to it as the factorization condition, satisfied by a much richer variety of couples of ``factor functions'' than just the ladder functions. In fact, this condition will be quite useful to construct the ladder functions. 

%Another issue has to be mentioned: we shall distinguish between ladder functions as functions satisfying the classical GHA (\ref{GHA1})-(\ref{GHA2}) in a purely algebraic sense (i.e. on some part of the phase space with an arbitrary function of $H$, $\alpha(H)$) and the dynamical objects for which $\alpha(H)$ is proportional to the (angular) frequency $\omega(H)$ of the system containing information about the dynamics of a particle under the influence of a given potential $V(x)$. 

In this paper, we assume that the potential $V(x)$ has the form of a well, and we want to describe the bounded motion of the system in that well. 
Then, the
motion with energy $H=E$ will be periodic between two turning points $x_\pm$ determined by two solutions of the equation $E=V(x)$ (where $p=0$).
The value of the physical frequency $\omega(H)$ (and its period $T(H)$) is given by
\begin{equation}
    \omega(H)\equiv \frac{2\pi}{T(H)} =\frac{2\pi}{\int_{x_-}^{x_+}\frac{ dx'}{\sqrt{H-V(x')}}} 
    \, . \label{frequency}
\end{equation}
Now, the ladder functions $A^\pm$ of this system, satisfying (\ref{GHA1}), will determine the constants of motion $Q^\pm$ (depending explicitly on time) defined by \cite{kuru12}
\begin{equation}
Q^\pm(x,p,t) = A^\pm(x,p) e^{\mp i\alpha(H)t} \,.
\end{equation}
This can be easily proved, since
\[
\frac{d Q^\pm}{d t} = \{Q^\pm ,H\} +\frac{\partial Q^\pm}{\partial t} = 0\,.
\]
The constant complex values  of $Q^\pm$ will be denoted by $q e^{\pm i\theta_0}$. Then,
the equations
\begin{equation}\label{constant}
A^\pm(x,p) e^{\mp i\alpha(H)t} = q e^{\pm i\theta_0}
\end{equation}
will lead to the motion $(x(t), p(t))$ of the system in the phase space in an algebraic way. As a consequence, the motion will have a frequency given by the
exponent $\alpha(H)$,
and therefore it should be equal to the physical frequency,  $\alpha(H)= \omega(H)$ for `fundamental' ladder functions $A^\pm$. 
Other ladder functions, such as $\tilde A^\pm = (A^\pm)^n$, $n=1,2,\dots$ will produce multiple frequencies 
$\alpha(H)= n\,\omega(H)$. The ladder operators (or functions) with multiple
frequencies $n>1$ have applications in some problems, for instance in order to find higher order symmetries of superintegrable systems \cite{calzada14,angel16}.

\subsection{Ladder functions in the variables $(x,H)$}

The object of this section is to change from the canonical variables $(x,p)$ to  another more practical set $(x,H)$ and to develop an integral formula for a ladder function satisfying  (\ref{GHA1}). This formula will not necessarily be useful to calculate an explicit (analytic) form for the ladder functions, but it will supply us with some information about its behaviour in the phase space.

If we introduce in the differential equation (\ref{GHA1}) the change of variables 
$(x,p)\rightarrow (x,H)$ with $p(x,H)=\pm \sqrt{H-V(x)}$ for each part of the half planes $p>0$ and $p<0$, then, using the chain rule, the Poisson bracket  (\ref{GHA1}) for the functions $A^\pm(x,H)$ becomes:
\begin{equation}
\{H, A^\pm(x,H)\}= -2p\,\frac{\partial A^\pm(x,H)}{\partial x}=\mp i\alpha(H) A^\pm(x,H)\, . \label{change of variable}
\end{equation}
%with the partial differentiation taken in the $(x,H)$ variables, i.e. H maintained constant. Indeed, we have
%\begin{equation}
%\frac{\partial A}{\partial p}=\frac{\partial A}{\partial H}\frac{\partial H}{\partial p}
%\end{equation}
%and
%\begin{equation}
%\frac{\partial A}{\partial x}=\frac{\partial A}{\partial x}|_H +\frac{\partial A}{\partial H}\frac{\partial H}{\partial x}.
%\end{equation}
Remark that this is not a change of canonical variables since $(x,H)$ are not canonical.
Notice also that eq.~(\ref{change of variable}) is only valid for each half-plane taken separately and not for $p=0$. 
This equation represents in fact two differential equations, one for each half-plane. 
To make this a bit clearer we introduce an index variable $\eta=\pm 1$ defined by $p=\eta \sqrt{H-V(x)}$. Then, eq.~(\ref{change of variable}) is rewritten as:
\begin{equation}
i\alpha(H) A^\pm(x,H,\eta)= \pm 2\eta\sqrt{H-V(x)}\, \frac{\partial A^\pm(x,H,\eta)}{\partial x}\,,  \label{differential eq 2}
\end{equation}
and its integration yields the formula (for each $H$, $V(x)<H$):
\begin{equation}
A^\pm(x,H,\eta)= B(H,\eta)\exp{\left( \pm i \eta \frac{\alpha(H)}{2}\int_{x_m}^{x} \frac{ dx'}{\sqrt{H-V(x')}}\right)},  \label{representation formula}
\end{equation}
where $B(H,\eta)$ is an integration constant, and the definite integral runs from any position $x_m\in(x_-,x_+)$ such that $V(x_m)<H$. 
%We can easily show (in Appendix 2) that this formula represents uniquely the solution of (\ref{differential eq 2}) in the two half-planes. 
Thus, formula (\ref{representation formula}) represents any algebraic ladder function $A$ that satisfies (\ref{GHA1}) with respect to the function $\alpha(H)$ in these two regions. 

The turning points $x_\pm$ are the ones for which  $p=0$, as mentioned above. In the coordinates $(x,H)$
they are given by $x_\pm(H)$ with $H = V(x_\pm)$, so that the points $x_\pm$
also determine the value $H(x_\pm)$.
The continuity of the ladder functions
$A^\pm(x,p)$ implies that  $A^\pm(x,H,\eta)$, defined by
(\ref{representation formula}), satisfy:
\[
\lim_{(x,H)\to(x_\pm,H(x_\pm))}A^\pm(x,H,+)= \lim_{(x,H)\to(x_\pm,H(x_\pm))}A^\pm(x,H,-) \,.
\]

\sect{Construction of ladder functions as a product of factor functions}

\subsection{Contribution of factor functions}
As it was mentioned in the introduction, we want to combine `factor functions' to obtain a ladder function. To do this we need a way to keep track of the contribution of each factor function. 
%In this section we will introduce an approach to this effect. 
 Let us take two factor functions $f(x,H)$ and $g(x,H)$, such that they verify the factorization
condition (\ref{Factorization}):
\begin{equation}\label{fact}
f^* f = \delta_f(H),\qquad g^* g = \delta_g(H) \, ,
\end{equation}
for some functions $\delta_f(H)$ and $\delta_g(H)$ of $H$.
Now, notice that since the Poisson bracket satisfies the Leibniz's rule, 
%which it shares with its quantum counterpart (the commutator):
\begin{equation}
 \{H,fg\}= f\{H,g\}+\{H,f\}g \, , \label{Leibniz}
\end{equation}
%This means that the $\Pi$ operator introduced in (\ref{operator pi}) acts like a derivative:
%\begin{equation}
% \Pi(fg) = \Pi(f)g+\Pi(g)f.
%\end{equation}
dividing by $fg$, we have
\begin{equation}
 \frac{\{H,fg\}}{fg}= \frac{\{H,f\}}{f}+\frac{\{H,g\}}{g}. \label{Pi-addition}
\end{equation}
 It is thus useful to introduce the following notation:
\begin{equation}
 \Lambda(f)= \frac{ \{H,f\}}{f}\, . \label{contribution operator}
\end{equation}
This function $\Lambda(f)$ will be called the `contribution' of $f$ and eq.~(\ref{Pi-addition}) will be reformulated as
\begin{equation}
 \Lambda(fg)= \Lambda(f)+\Lambda(g). \label{contribution add}
\end{equation}
The ladder bracket (\ref{GHA1}) is now equivalent to
 \begin{equation}
 \Lambda(A^\pm)= \mp i\alpha(H). \label{lambda yields alpha}
\end{equation}
Hence, if after adding the contributions (\ref{contribution add}) of two factor functions, the sum is a function depending only on $H$, then the product $f g$ will constitute a good ladder function (at least algebraically speaking).

%At this point, nothing is really gained from this formulation of the problem. Its usefulness comes from noticing that the contribution of some types of functions are easier to handle.% 
The most useful ansatz for a factor function in the search of ladder functions, suggested by the form of other simpler cases \cite{kuru08}, is
 \begin{equation}\label{ff}
 f(x,H)=a(x,H) + i\, b(x,H) p \,,
 \end{equation} 
with $a(x,H)$ and $b(x,H)$ real functions depending on the variables $x$ and $H$.  %and $x$-note:the formula below doesn't depend on b=b(x)%,
For such a function, using $p^2= H-V(x)$ and $p'=-\frac{V'(x)}{2p}$, its contribution is given 
%(assuming that it satisfies the factorization condition (\ref{Factorization})) 
by:
\begin{equation}
\Lambda(f(x,p))=  i\,\frac{2(H-V)(a'b-b'a)+V'b\,a}{\delta_f(H)}\, , \label{contribution a+ibp}
\end{equation}
where the derivative is taken with respect to $x$ maintaining $H$ constant.

Another useful fact about $\Lambda$ is that we can handle exponents more easily. Let us assume that we can define the exponent of a function $f$ using a well chosen determination of the complex logarithm $f^{\rho(H)}=\exp[\rho(H)\log(f)]$, then it is easily seen from (\ref{Leibniz}) and  (\ref{contribution operator}) that we will have:
 \begin{equation}
\Lambda(f^{\rho(H)})= \rho(H)\Lambda(f). \label{exponent}
\end{equation}
Some technicalities arising from the proper definition of exponent in the complex plane will have to be addressed if necessary.
%later to deal with the full extent of our approach%.

\subsection{Signature}

%{\color{blue} the minimal assumption here is differentiability.  The poisson bracket definition implies partial differentiability, and there might be a way to extract the continuity of both partial derivatives from all the poisson brackets and continuity of A and alpha's. In practice, these issues are not that relevant(we are dealing with smooth functions most of the time). But i don't want to state something false so i modified the text here.}
%
%As can be understood from (\ref{representation formula}), it is not enough to have an algebraically valid ($\alpha(H)$ arbitrary) GHA defined in some part of the plane (say a region disconnected by the line $p=0$). We want also obtain a dynamically valid ladder function 
%. It is not even obvious (though it may be suspected) that the requirement that the function be continously differentiable at $p=0$ will be enough (without introducing dynamical considerations). 
In this subsection we will introduce a necessary and sufficient condition for a ladder function 
to be dynamically valid, i.e., $\alpha(H)=n\omega(H)$, $n=1,2,\dots$ (under the standard assumption of continuous differentiability). 
%As it turns out, making our ladder functions satisfy the  condition also completely or partially solves the problem caused by the singularity and discontinuity problems at $p=0$ in (\ref{representation formula}).
We start by assuming that we have the functions $A^\pm(x,p)$ which satisfy (\ref{GHA1}) with a certain function $\alpha=\alpha(H)$ in a symmetric simply connected region $R$ of the plane (including a segment of $p=0$ as its axis of symmetry%, which depend on the chain rule
). We can then introduce two corresponding functions  $A^\pm(x,H,\eta)$ of 
$(x,H)$ in each region $p\geq 0$ and $0\leq p$, using the index parameter $\eta=\pm 1$ defined in Sect.~2 ($p=\eta\sqrt{H-V(x)})$. Now, this means that for the two half-regions $p>0$ and $p<0$, the representation (\ref{representation formula}) is valid. 
%The differentiability of $f(x,p)$ on the whole region $R$ then implies:
%\begin{equation}
%  f_{\eta}(x_\pm,H)=  B(H,\eta)
%  \exp{\left(-i\eta\frac{\alpha(H)}{2}\int_{x_m}^{x_\pm}\frac{ dx'}{\sqrt{H-V(x')}}\right)}\,,
%\end{equation}
%with $x_{\pm}$ the turning points where $p=0$.
Then, if $\alpha(H)=n\omega(H)$ with $\omega(H)$ defined by (\ref{frequency}), we will have that
\begin{equation}
    A^\pm(x_+,H,\eta)=(-1)^n A^\pm(x_-,H,\eta). \label{antiperiodicity}
\end{equation}
To see this, we just divide $A^\pm(x_+,H,\eta)$ by $A^\pm(x_-,H,\eta)$ and we expand the argument of the exponential using the fact that $\int_{x_-}^{x_m}+\int_{x_m}^{x_+}=\int_{x_-}^{x_+}$. Taking into account (\ref{frequency}), this gives $e^{in\pi}=(-1)^n$. In the same way, it is easily shown that (\ref{antiperiodicity}) implies $\alpha(H)=n\omega(H)$ (sufficiency). We call condition (\ref{antiperiodicity}) antiperiodic if $n$ is odd (this contains the privileged case $n=1$, i.e. the `fundamental' ladder functions) and periodic if $n$ is even. 

Now, we can define for any factor function $f(x,H)$ its signature 
$\Gamma(f)$  by
\begin{equation}
    \Gamma(f)= \frac{f(x_+,H)}{f(x_-,H)} \label{signature} \,.
\end{equation}
Thus, $\Gamma(f)=1$ corresponds to the periodic case and $\Gamma(f)=-1$ to the antiperiodic case. The following simple property of the signature is really useful:
\begin{equation}
    \Gamma(fg)= \Gamma(f)\Gamma(g). \label{signature product}
\end{equation}
If $A^\pm= fg$, this will allow us to calculate the signature of $A^\pm$ from that of each factor function. 

The behaviour of the signature with respect to exponentiation is subtler and has to be examined more carefully.
  First, remark that in the case of a function of the form $f$ given by (\ref{ff}), the signature reduces to the calculation of the ratio:
 \begin{equation}
    \Gamma(f)= \frac{a(x_+,H)}{a(x_-,H)}\, . \label{signature a+ibp}
\end{equation}
Therefore, if both $a(x_+,H)$ and $a(x_-,H)$ are positive, we will simply have:
 \begin{equation}
    \Gamma(f^{\rho(H)})= \Gamma(f)^{\rho(H)}. \label{signature exponent}
\end{equation}
%However in general, this is not the case and we will have to multiply (\ref{signature exponent}) by a correction factor calculated from the (chosen) definition of exponents in the complex plane. For the next two sections, Equation (\ref{signature exponent}) will be shown to be sufficient. 

Finally, the function $f$ given by (\ref{ff}) that satisfies the factorization condition (\ref{Factorization}) with its complex conjugate, leads to
\begin{equation}
    a^2+b^2p^2= \delta_f(H).
\end{equation}
From this and (\ref{signature a+ibp}), it is easy to see that $\Gamma(f)^2=1$ so that $\Gamma(f)=\pm 1$ at $p=0$. This will allow us to determine the signature of the function by simple analytic %analysis 
arguments. 

\sect{The Rosen-Morse II system}

\begin{figure}
\centering
\includegraphics[width=0.45\textwidth]{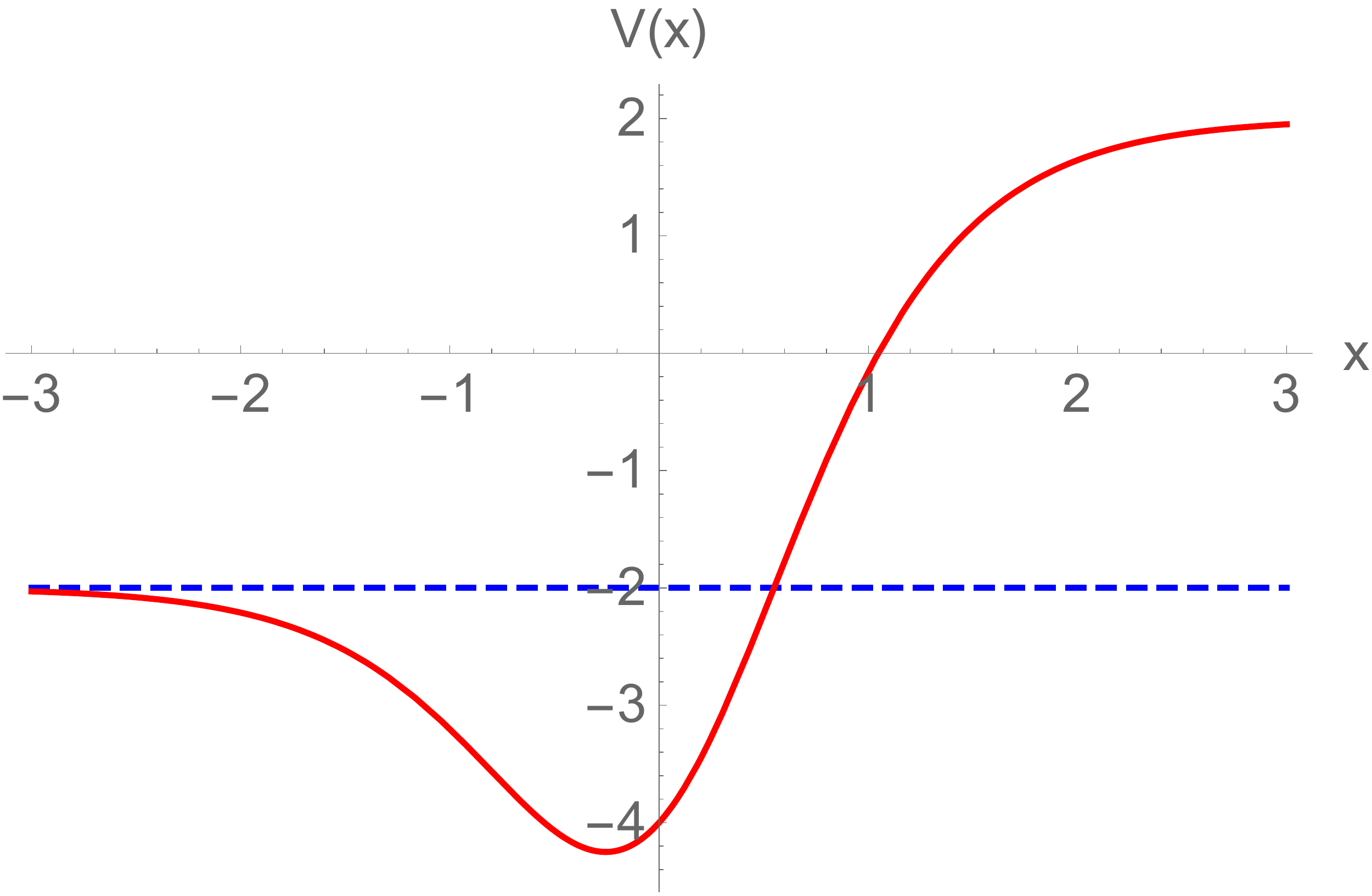}
\caption{\small Plot of the Rosen-Morse II potential for $B=2$, $C=4$.
The dashed line corresponds to $E=-B$.
\label{fig1}}
\end{figure}

In this section we consider the RMII system \cite{cooper00,gadella17} defined by the Hamiltonian
\begin{equation}\label{hrm}
    H= p^2+B\tanh x-\frac{C}{\cosh^2 x}\,, \qquad x\in \mathbb R\, ,
\end{equation}
where $B$ and  $C$ are real parameters. In order to have the shape of a well, we will
assume that $C>0$. We will look for ladder functions of this system in the
region of the phase-space $H(x,p)< -|B|$, where the motion is bounded and periodic
(see Fig.~\ref{fig1}).

%%%%%%%%%%%%%
%%%%%%%%%%%%%%

%We have computed the frequency corresponding to such bounded
%motions of the RMII system by means of the integral (\ref{frequency}); the resulting explicit formula is
%\begin{equation}\label{omegaRM}
%\omega(H)=  \frac{4\sqrt{H^2-B^2}}{\sqrt{-H+B}+\sqrt{-H-B}}\,.
%\end{equation} 

\subsection{Ladder functions}

We will look for the fundamental ladder functions satisfying (\ref{GHA1}) and such that $\alpha(H)$ is the frequency. The process is similar to the simplest cases
as the harmonic oscillator, P\"oschl-Teller, etc. but in this case it is a bit more
complex. We will need two sets of factor functions that factorize the RMII Hamiltonian in the sense (\ref{Factorization}), such that
they will be the basic ingredients of the fundamental ladder functions. 
%We thus introduce four different factor functions $f_{\epsilon}, \ g_{\epsilon}$, where $\epsilon=\pm 1$. 

The first set of factor functions $f_\epsilon$, where $\epsilon=\pm 1$, is given by
\begin{equation}
    f_{\epsilon} = a_{\epsilon}+ ib\, p,  \label{definition f RM 1}
\end{equation}
where
\begin{equation}
   b(x)= \cosh x, \quad a_{\epsilon}(x,H)= b(x)\left( \phi_{\epsilon}(H)\tanh x+\frac{B}{2 \phi_{\epsilon}(H)}\right). \label{definition a RN 1}
\end{equation}
Then, the factorization condition (\ref{fact}) is  fulfilled  with 
\begin{equation}
    \delta_{f_{\epsilon}}(H) = C-\phi_{\epsilon}(H)^2, \label{Factorizationf}
\end{equation}
where  $\phi_{\epsilon}(H)$ is 
\begin{equation}\label{phie}
 \phi_{\epsilon}(H)= \frac{\sqrt{-H+B}+\epsilon \sqrt{-H-B}}{2}\,,
\end{equation}
and
\begin{equation}
   \phi_{\epsilon}(H)^2 = \frac{-H+\epsilon\sqrt{H^2-B^2}}{2}\, . \label{definition of Phi RM 1}
\end{equation}

The second set of factor functions $g_\epsilon$ is given by
\begin{equation}
    g_{\epsilon} = c_{\epsilon}- id_{\epsilon} p, 
    \label{definition g RM 1}
\end{equation}
%{\color{blue} I modify the def of g(complex conjugation just reverses the sign of the contribution), so that $\lambda(g)=-i\sqrt{-H+\epsilon B}(...)$}
where
\begin{equation}\label{second}
d_{\epsilon}(x)= \frac{1}{\tanh x-\epsilon}, \quad c_{\epsilon}(x,H)= 
d_{\epsilon}(x)\left(\frac{(B+2 \epsilon C)\tanh x +\epsilon B-2 (H+ C)}{2\sqrt{-H+\epsilon B}}\right).
\end{equation}
%{\color{red} Vero: Check again the expression of c.}
The factorization condition (\ref{fact}) is satisfied with the function 
\begin{equation}
    \delta_{g_{\epsilon}}(H) =  \frac{B^2+4C(C+H)}{4(-H+\epsilon B)}\, .\label{Factorizationg}
\end{equation}

One may be surprised by the fact that there are essentially two
simple forms of ladder functions given by $f_\epsilon$
and $g_\epsilon$. However, the origin of these
solutions is easy to explain. The first factorization comes after writing 
the Hamiltonian (4.1) in the form
\begin{equation}
C=  (B\tanh x-H)\cosh^2 x+p^2\cosh^2 x= 
(a_\epsilon+i b p)(a_\epsilon-i b p) +\phi_\epsilon(H)^2 \,.
\end{equation}
The second factorization comes if we rewrite (4.1) in terms
of $\tanh x$,
\begin{equation}\label{hrm2}
    H+2C= p^2+(B+\epsilon 2C)\tanh x+C(\tanh x -\epsilon)^2\,, \qquad x\in \mathbb R\, ,
\end{equation}
or
\begin{equation}\label{hrm2}
C= \frac{p^2}{(\tanh x -\epsilon)^2}+\frac{(B+\epsilon 2C)\tanh x-(H+2C)}{(\tanh x -\epsilon)^2}
\,, \qquad x\in \mathbb R\, ,
\end{equation}
whose immediate factorization leads to the second pair (\ref{definition g RM 1})-(\ref{second}).

%{\color{blue} i added here further justification for the introduction of the factor functions, see if you agree with this}
%We note that these functions were chosen so as to make their contribution $\Lambda(f_{\pm 1}),\Lambda(g_{\pm 1})$ affine($=q(H)\tanh x+w(H)$) in the variable $\tanh x$, using (\ref{contribution a+ibp}) and noticing that the potential $V(x)$ can be written in terms this variable as $V(x)=B\tanh x-C(1-\tanh^2 x)$. The reason for this choice will soon become clear. 

The calculations of the contribution and signature of each function are now straightforward. In the Appendix we give an analytic argument for the values of the signature. The results are given in the following set of relations:
\begin{equation}\label{rmlg}
\begin{array}{ll}
\displaystyle  \Lambda(f_{\epsilon})= i(2 \phi_{\epsilon}(H) +\frac{B}{\phi_{\epsilon}(H)}\tanh x),
  \qquad & 
  \Gamma(f_{\epsilon})=-\epsilon,
\\ [2.5ex]
\displaystyle
\Lambda(g_{\epsilon})= - 2 i%\epsilon%  
\sqrt{-H+\epsilon B}\,(\tanh x +\epsilon),
	\quad &
 	\Gamma(g_{\epsilon})=-1\,. 
\end{array}
\end{equation}
%{\color{blue} i added - to the contribution of g, bcs the new g's are complex conjugate of the old}
%{\color{red} Vero: Check again $\Lambda(g_{\epsilon})$ }

Our strategy is as follows: we form different products of two of these factor functions such as $f_{\epsilon_1}^{\gamma(H)}g_{\epsilon_2}^{\sigma(H)},\  f_{1}^{\gamma(H)}f_{-1}^{\sigma(H)},\ g_{1}^{\gamma(H)}g_{-1}^{\sigma(H)}$ and choose the exponents to make the $\tanh x$ dependence in (\ref{rmlg}) cancel against each other  and, at the same time, to satisfy the antiperiodic condition. For all these products we have to carefully treat the exponents which could take values in the complex plane in general. There are however products that we can deal with without worrying too much about this issue. Indeed, as we have shown in the derivation of the signature of $f_{\pm 1}$ (see Appendix), we have the useful result $a_{-1}(x,H)>0$. This means that the image of $f_{-1}=a_{-1}+ibp$ is confined to the half-plane ${\rm Re}(z) >0$ of the complex plane. In this region, we can simply use the principal determination of the logarithm (which is continuous everywhere except on the half-line  $\{z : {\rm Re}(z)\leq 0, {\rm Im}(z)=0  \}$ ) to define the exponent  $f_{-1}^{\gamma(H)}$ as a continuous function of $(x,p)$ (in the region where $H<-B$). Also this means that we can use (\ref{signature exponent}) to calculate its signature: $\Gamma(f_{-1}^{\gamma(H)})=1$. In order to get a ladder function, let us thus define the following products:
\begin{equation}\label{ladderf}
    A_{\epsilon}= f_{-1}^{\gamma_{\epsilon}(H)}g_{\epsilon}
\end{equation}
with
\begin{equation}
   \label{gamma RM 2} \gamma_{\epsilon}(H)= \frac{2 %\epsilon% 
    \phi_{-1}(H)\sqrt{-H+\epsilon B}}{B}\,.
\end{equation}
%{\color{blue} Lucas: i modified the exponent to go along with the redefinition of the g's}
We use (\ref{contribution add}) and (\ref{exponent}) to deduce the
formula
\begin{equation}
  \label{RMalpha=nomega}  \Lambda(A_{\epsilon})= \gamma_{\epsilon}(H)\Lambda(f_{-1}) + \Lambda(g_\epsilon)=- i \epsilon\,\frac{4\sqrt{H^2-B^2}}{\sqrt{-H+B}+ \sqrt{-H-B}}
    \equiv -i \epsilon \alpha(H)\,. 
\end{equation}

%{\color{red} Vero: Check again $\Lambda(A_{\epsilon})$ and  $ \gamma_{\epsilon}(H)$. It is important because it will select the ladder plus or minus.}{\color{blue} Lucas: J'ai calculÃ© les lambdas et il y avait une erreur avec la nouvelle dÃ©finition de g, (c) il fallait enlever les epsilons dans (\ref{rmlg}) et de maniÃ¨re correspondante dans la definition de gamma. Si on fait Ã§a, j'ai vÃ©rifiÃ© que le reste est O.K. Mais j'ai changÃ© la def des g pour pouvoir rendre l'exposant gamma positif, tout Ã§a ajoute des - dans les contributions de g et de A(Ã§a correspond aussi au choix de K et N}

%Thus, we have an algebraically valid ladder function on all of the region $H<-|B|$ with the equality $\alpha(H) = n\omega(H)$.% %Indeed%, 
Now, the properties of the signature (\ref{signature product}) and (\ref{signature exponent})
together with (\ref{rmlg}), applied for the chosen factor functions in
(\ref{ladderf}), allows us to check that
$
    \Gamma(A_{\epsilon})=-1
$.
Thus, we conclude that we have an algebraically valid ladder function for the region $H<-|B|$ with the equality $\alpha(H) = n\omega(H)$, where $n$ must be odd. In the next subsection we show that $n=1$ by considering the case $B=0$. This means that we have found a set of fundamental ladder functions given by $A^{\pm}=A_{\pm 1}$. In fact, we have independently checked that the value of the natural frequency is indeed given in (\ref{RMalpha=nomega}) by a direct evaluation of the integral (\ref{frequency}) for
the RMII potential: 
\begin{equation}
    \omega(H) =\frac{2\pi}{\int_{x_-}^{x_+}\frac{ dx'}{\sqrt{H-V(x')}}} 
    = \frac{4\sqrt{H^2-B^2}}{\sqrt{-H+B}+ \sqrt{-H-B}}
    \, . \label{frequency2}
\end{equation}

\begin{figure}
\centering
\includegraphics[width=0.3\textwidth]{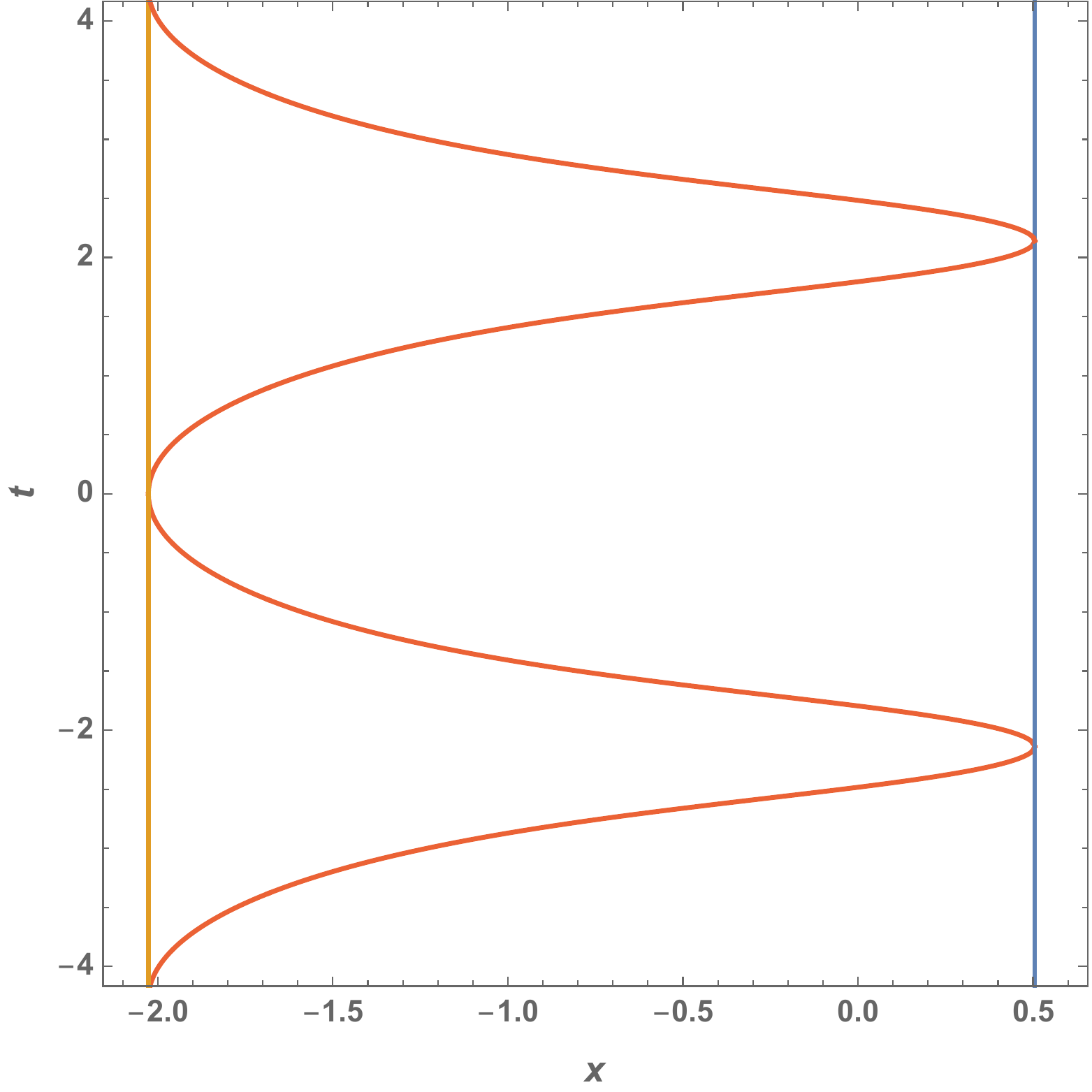}
\includegraphics[width=0.3\textwidth]{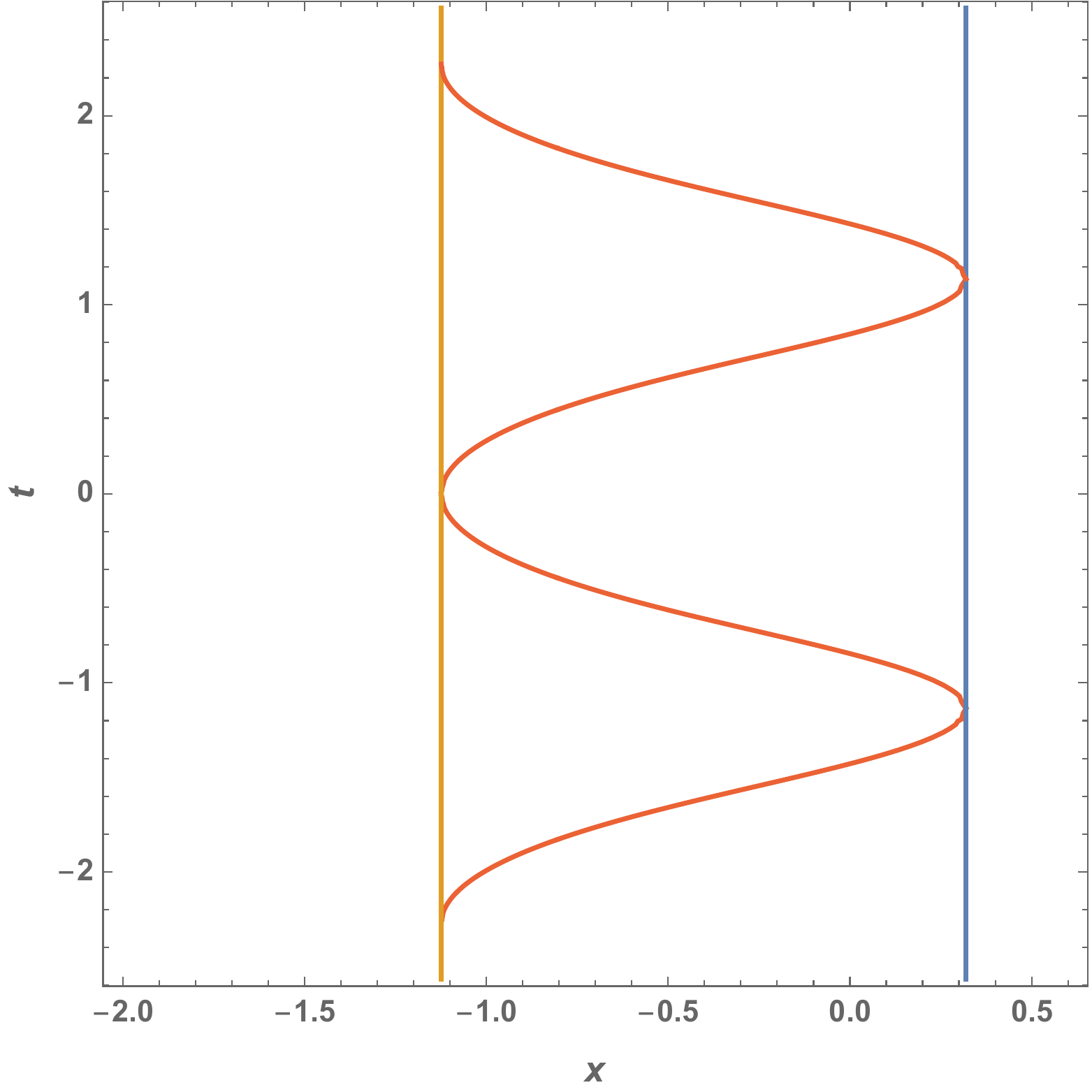}
\includegraphics[width=0.3\textwidth]{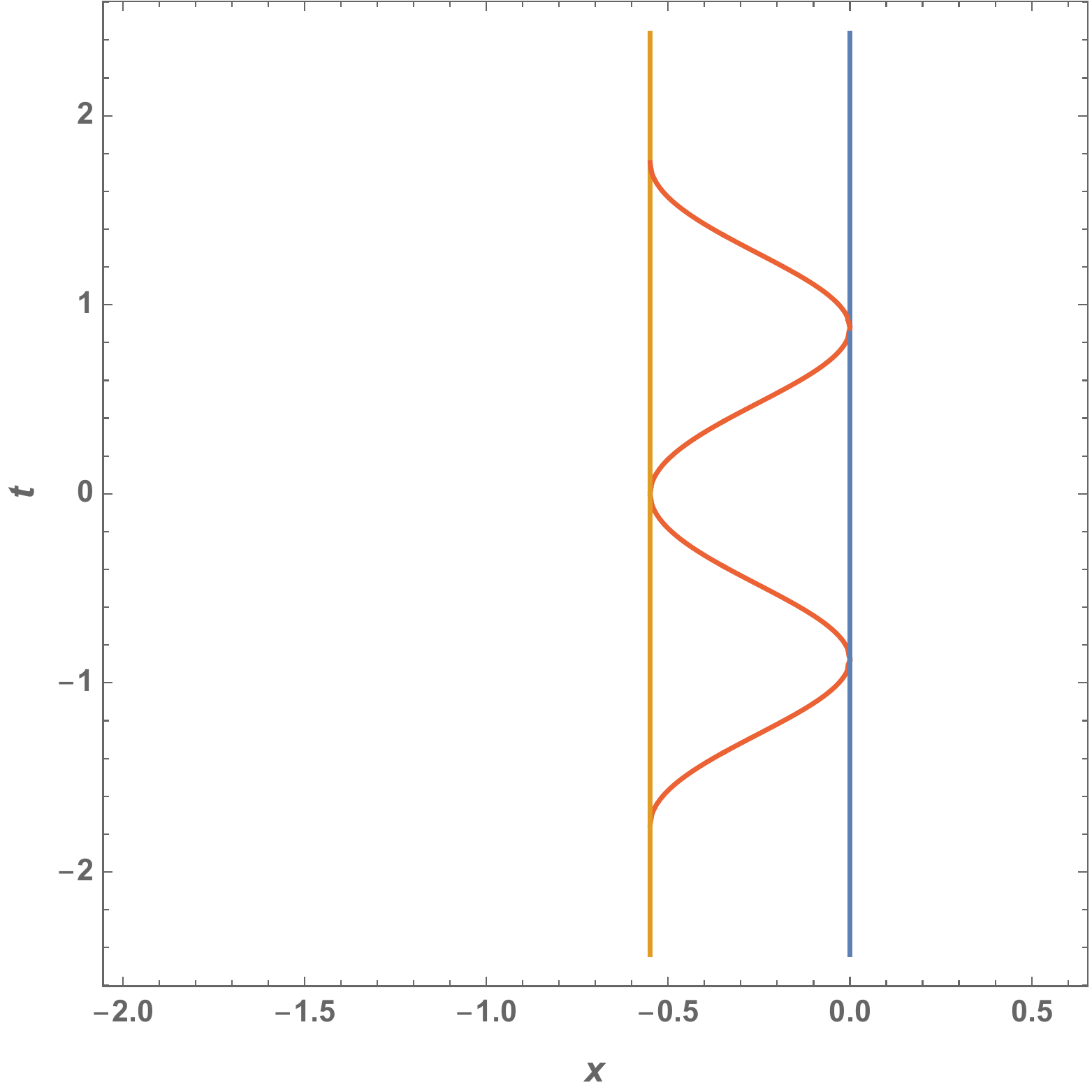}
\caption{\small Plot of the motion $x(t)$ in the Rosen-Morse II potential with $B=2$, $C=4$,
and energy values:  $E=-2.2$ (left), $E=-3$ (center) $E=-4$ (right). The left and right vertical lines are for the values $x_-$ and $x_+$ bounding the
motion. The motion in the graphic is restricted to two oscillations. \label{fig2}}
\end{figure}

As a consequence, according to the expression of the constant of motion (\ref{constant}) and (\ref{ladderf}),
the motion of a particle in the MRII potential (for $B>0$ and $\epsilon=-1$) is given by 
\begin{equation}\label{motion1}
\begin{array}{l}
\displaystyle \gamma_{-1}(E)\,{\rm arctan}\left[\frac{p(x,E)}{\phi_{-1}(E)\tanh x+b/(2\phi_{-1}(E)) }\right]
+ 
\\[3.ex]
\qquad \qquad\qquad
\displaystyle {\rm arctan}\left[\frac{-2p(x,E)\sqrt{-E-B}}{(B-2C) \tanh x- B -2(E+C)}  \right]
-\omega(E) t = \theta_0 \, ,
\end{array}
\end{equation}
where $\theta_0$ is a constant fixing the initial time. A plot of some examples of this motion for some particular values of the energy $E$ is shown in Fig.~\ref{fig2}.

%%%%%%%%%%%%%
%%%%%%%%%%%%%%

\subsection{The P\"oschl-Teller potential}

The (hyperbolic) P\"oschl-Teller (PT) potential is a particular case of the Rosen-Morse potential when $B=0$: 
\begin{equation}
    H=p^2-\frac{C}{\cosh^2 x}\,, \qquad x\in \mathbb R.
\end{equation}

In this section, we want to check that the ladder functions of the RMII potentials in the previous subsection, when taking the limit $B\to 0$, agree with the known ladder functions of PT \cite{kuru08,kuru07}. In this way, we will also prove the claim made at the end of the subsection that we have fundamental ladder functions for our potentials ($n=1$ in (\ref{RMalpha=nomega})).
The realization of the GHA of this system is considered at the quantum and classical levels in \cite{kuru07,kuru09}. We quote the classical ladder functions obtained there:
\begin{equation}
    A^{\pm}_{PT}(x,p)=\mp ip\cosh x  +\sqrt{-H} \sinh x \,. \label{PT}
\end{equation}
%We note that this agrees with the expression of the factor functions $f_{1},(f_{1})*$ from the previous subsection in the limit $B\rightarrow0$.
%The frequency of the system can be obtained from $\frac{\Lambda}{i}$ in the manner prescribed by ($\ref{lambda yields alpha})$, knowing that it is a fundamental ladder function ($\alpha(H)=\omega(H)$). We use (\ref{contribution a+ibp}) to deduce that:
The value of the function $\alpha(H)$ obtained in that reference is
\begin{equation}
    \alpha_{PT}(H)=\omega_{PT}(H)=2\sqrt{-H}.
\end{equation}
In this case, the functions (\ref{PT}) constitute a system of fundamental ladder functions.

 We now show that the expression given in the previous section for the ladder functions 
 \[
 A^{\pm}\equiv A_{\pm 1}=f_{-1}^{\gamma_{\pm 1}}g_{\pm 1}
 \] 
 agrees with (\ref{PT}) in the limit  $B\rightarrow0$.

Firstly, after some computations, we can write the expression for $f_{\pm 1}$ as follows
\begin{equation}\label{ffpt}
    f_{\pm 1}=  \phi_{\mp 1} \cosh x+ \phi_{\pm 1}\sinh x +i p\cosh x\,.
\end{equation}
We deduce from this and from (\ref{phie}) that for $B=0$ ($\phi_{-1}=0$, $\phi_{+1}=\sqrt{-H}$), 
\begin{equation}
    f_{-1}(x,B=0)=(\sqrt{-H}+ip)\cosh x    \,.
\end{equation}
We then write the expression for $g_{\epsilon}$ for $B=0$ from (\ref{definition g RM 1}) as
\begin{equation}
    g_{\epsilon}(x,B=0)= \frac{\epsilon\,C}{\sqrt{-H}}+\frac{\sqrt{-H}-ip}{\tanh x -\epsilon}\,.
\end{equation}
The value of the exponent (\ref{gamma RM 2}) is then seen to be
\begin{equation}
    \gamma_{\epsilon}(B=0)=\frac{\sqrt{-H}}{\phi_{+1}(B=0)}=1 \,.
\end{equation}
So, finally the calculation of $A^{\pm}$ reduces to:
\begin{equation}
     A_{\epsilon}= f_{-1}(B=0)g_{\epsilon}(B=0)= -\frac{C}{\sqrt{-H}}(\sqrt{-H}\sinh x-i\epsilon p\cosh x)
     \,.
\end{equation}
 We conclude from this and (\ref{PT}) that indeed our expressions in the limit $B=0$ corresponds to the known results (\ref{PT}) up to a function of $H$:
 \begin{equation}
     A^{\pm}(B=0)=-\frac{C}{\sqrt{-H}}A^{\pm}_{PT}.
 \end{equation}
 Thus, we have
 \begin{equation}
     \alpha_{RM}(B=0)=n\omega(B=0)=\omega_{PT}=2\sqrt{-H}.
 \end{equation}
From this, we deduce that $n=1$ for RMII. Remark that, in the 
simpler case of the PT system, the ladder functions (\ref{PT})
can be obtained simply in the form $f_{+1}(x,p) = A^-(x,p)$, making
use of one factor function (\ref{ffpt}).

%This together with the continuity properties of $A_{\epsilon}$ and the considerations of the last section, allow us to conclude that (up to an undetermined $n= 1,  2,...$):
%
%\begin{equation}
%\alpha(H)= n\omega(H)=  \frac{4\sqrt{H^2-B^2}}{\sqrt{-H+B}+\sqrt{-H-B}}.
%\end{equation}
%
%We have computed the frequency of the Rosen-Morse II system by the action-angle
%method and this equality is satisfied for $n=1$, so that $\alpha(H) = \omega(H)$
%and therefore, in fact, we have obtained  a set of fundamental ladder functions.

%Thus we have a dynamically valid ladder function because $\alpha(H)= -n\epsilon\omega(H)$. 

\sect{The curved Kepler-Coulomb system}

The Hamiltonian associated to the curved Kepler-Coulomb problem in the radial coordinate $r$  can be written as \cite{Stevenson41,Infeld45,ranada99,Schrodinger40,ballesteros06,
Quesne16}
(sometimes this is also referred to as the Eckart 
\cite{Eck} or Hulth\'en \cite{Hul,Flu} potential):
 \begin{equation}\label{kch}
    H_\kappa=p^2 + V_\kappa(r)= p^2-\frac{B \sqrt{\kappa}}{\tanh \sqrt{\kappa} \, r}+\frac{\ell^2 \kappa}{\sinh^2\sqrt{\kappa}\, r}, %\qquad 0<r<+\infty \,,
\end{equation}
where $\ell$ is the angular momentum and $\kappa$ is a curvature parameter. Depending on the sign of $\kappa$ this expression will include the KC problem on the sphere
($\kappa<0$), the plane ($\kappa=0$), or hyperboloid ($\kappa>0$). The corresponding formulas 
for the values $\kappa = -1,\, \kappa =0,\, \kappa =1$, are respectively
\begin{equation}\label{cases}
\begin{array}{ll}
\displaystyle H_1= p^2-\frac{B  }{\tanh   \, r}+\frac{\ell^2  }{\sinh^2 \, r}, \qquad &0<r<+\infty \,,
\\[2.ex]
\displaystyle H_0= p^2-\frac{B  }{ r}+\frac{\ell^2  }{ r^2}, \qquad &0<r<+\infty \,,
\\[2.ex]
\displaystyle H_{-1}= p^2-\frac{B  }{\tan   \, r}+\frac{\ell^2  }{\sin^2 \, r}, \qquad &0<r<\pi \,.
\end{array}
\end{equation}
In Fig.~\ref{fig2} it is shown the form of the potentials for different values of $\kappa$.
In order to have the shape of a well for $\kappa\geq0$, we will assume that the potential is attractive $(B>0)$ and
\[
 2 \ell^2 \sqrt{\kappa}<B ,\qquad 0\leq\kappa  \,.
\]
Unless otherwise stated, we will restrict in this section to the case $\kappa>0$. We also remark that from the curved KC Hamiltonian $H_1$ in (\ref{cases}) we can formally get the RMII Hamiltonian $H_{\rm RM}$ by means of 
a complex displacement:
\begin{equation}\label{rmkc}
H_{\rm RM}(p,x) = H_1(p,x +i \frac{\pi}2)\,.
\end{equation}

%%%%%%%%%%%%%
\begin{figure}
\centering
\includegraphics[width=0.5\textwidth]{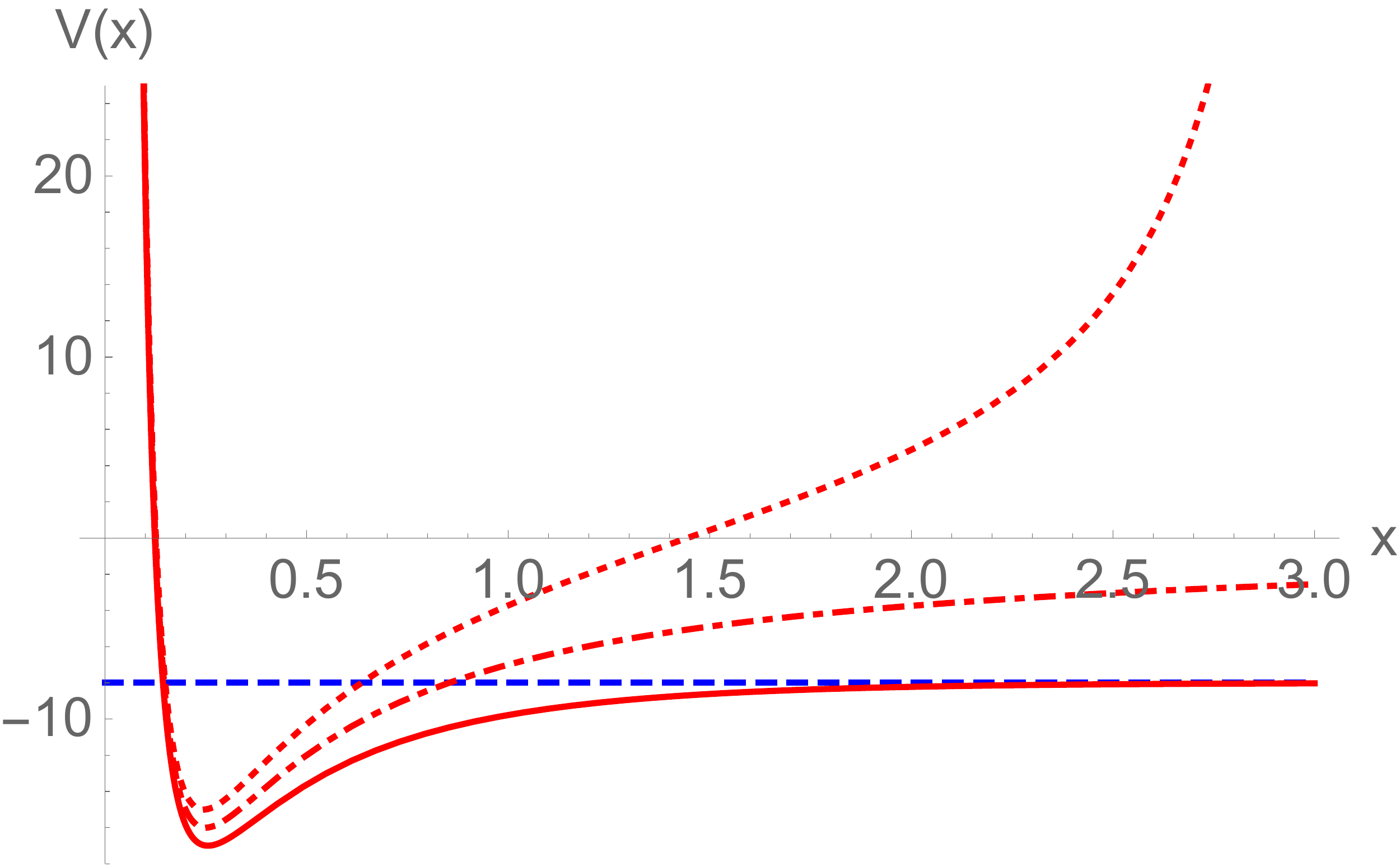}
\caption{\small Plot of the curved Kepler-Coulomb potential with the parameters $B=8$, $\ell^2=1$ for
$\kappa=1$ (continuous), $\kappa=0$ (dashed-dotted), $\kappa=-1$ (dotted).
The dashed blue horizontal line corresponds to $E=-B$.
\label{fig3}}
\end{figure}
%%%%%%%%%%%%%%

We computed directly the physical frequency $\omega(H)$ for arbitrary $\kappa$, and the
explicit formula is given by
\begin{equation}\label{omega}
\omega(H)=\frac{4\sqrt{\kappa}\sqrt{H^2-B^2\kappa}}{\sqrt{-H+B\sqrt{\kappa}}-\sqrt{-H-B\sqrt{\kappa}}}\,.
\end{equation}
Therefore, we would like to find ladder functions with associated bracket function
$\alpha(H) = \omega(H)$ as given by (\ref{omega}).
As before, following our procedure, we introduce the four factor functions $f_{\epsilon}$ and $\ g_{\epsilon}$ with $\epsilon =\pm1$. The 
 first factor function $f_{\epsilon}$ is defined as in (\ref{definition f RM 1}),
but now
\begin{equation}
   b(r)=  {\sinh \sqrt{\kappa}\,r} , \quad a_{\epsilon}(r,H)= b(r)\left( \frac{\tilde{\phi}_{\epsilon}(H)}{\tanh \sqrt{\kappa}\, r}-\frac{B \sqrt{\kappa}}{2 \tilde{\phi}_{\epsilon}(H)}\right), \label{definition a KC 1}
\end{equation}
with
\begin{equation}
   \tilde{\phi}_{\epsilon}^2 = \frac{-H+\epsilon\sqrt{H^2-B^2 \kappa}}{2}. \label{definition of Phi KC 1}
\end{equation}
The factor function $g_\epsilon$ is defined by the formula (\ref{definition g RM 1}), where
\begin{equation}
d_{\epsilon}(r)= \frac{-1}{{\sqrt{\kappa}} (\coth {\sqrt{\kappa}}\ r +\epsilon)}, 
\end{equation}
\begin{equation}
c_{\epsilon}(r,H)=d_{\epsilon}(r)\left( \frac{-( B {\sqrt{\kappa}}+\epsilon 2\ell^2 \kappa)  \coth {\sqrt{\kappa}}\ r + \epsilon B{\sqrt{\kappa}}-2(H+\ell^2 \kappa)}{2 \sqrt{-H+\epsilon B {\sqrt{\kappa}}}}\right). 
\end{equation}

In this case, the factorization condition (\ref{fact}) is satisfied, respectively, with 
\begin{equation}
    \delta_{f_{\epsilon}}(H) = -\ell^2 \label{delta f kc} \kappa+\tilde{\phi}_{\epsilon}(H)^2, 
\end{equation}
and
\begin{equation}
    \delta_{g_{\epsilon}}(H) = \frac{B^2\kappa +4l^2(l^2 \kappa+H)}{4(-H+ \epsilon B {\sqrt{\kappa}})}.\label{Factorizationg}
\end{equation}
At this point, we remark the similarity of the previous formulas of the factor functions for KC for $\kappa=1$, with those for the Rosen-Morse II systems, in agreement with the above comment (\ref{rmkc}) on the relation between these Hamiltonians.

The calculation of the contribution and signature proceed in exactly the same way as in the previous section and we obtain:
\begin{equation}\label{signature g}% \label{contribution g}
\begin{array}{ll}
\displaystyle    \Lambda(f_{\epsilon})=  i\sqrt{\kappa}(2{\tilde\phi}_{\epsilon}(H) -\frac{B \sqrt{\kappa}}{\tilde{\phi}_{\epsilon}(H)}\coth \sqrt{\kappa}\ r),
    \quad &  \Gamma(f_{\epsilon})=\epsilon,
\\  [2.5ex]  
\Lambda(g_{\epsilon})=  2i\sqrt{\kappa} \sqrt{-H+ \epsilon B{\sqrt{\kappa}}} \ 
    (\epsilon -\coth {\sqrt{\kappa}}\ r),
\quad & \Gamma(g_{\epsilon})=-1 \,.
\end{array}    
\end{equation}

\begin{figure}
\centering
\includegraphics[width=0.3\textwidth]{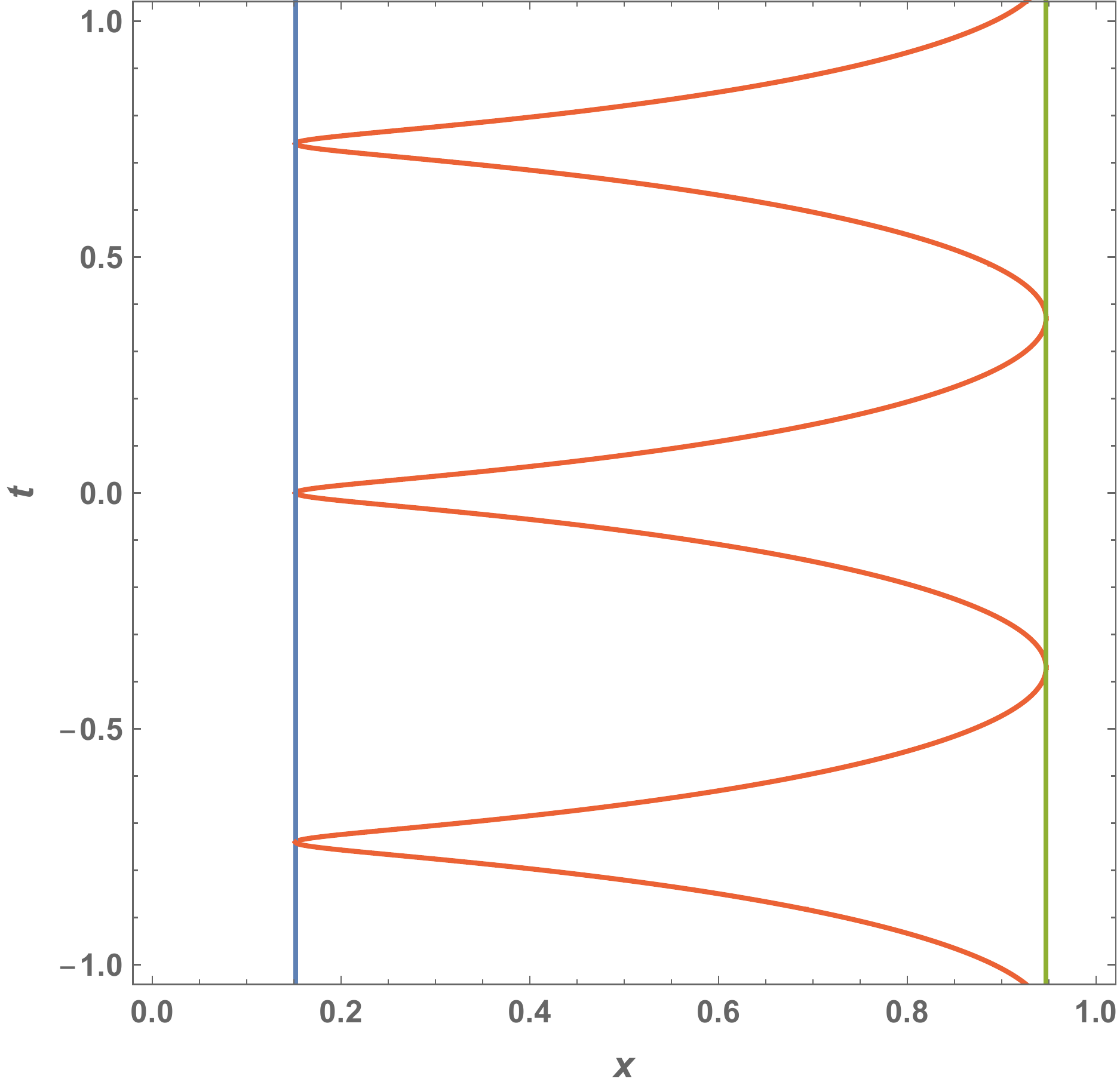}
\includegraphics[width=0.3\textwidth]{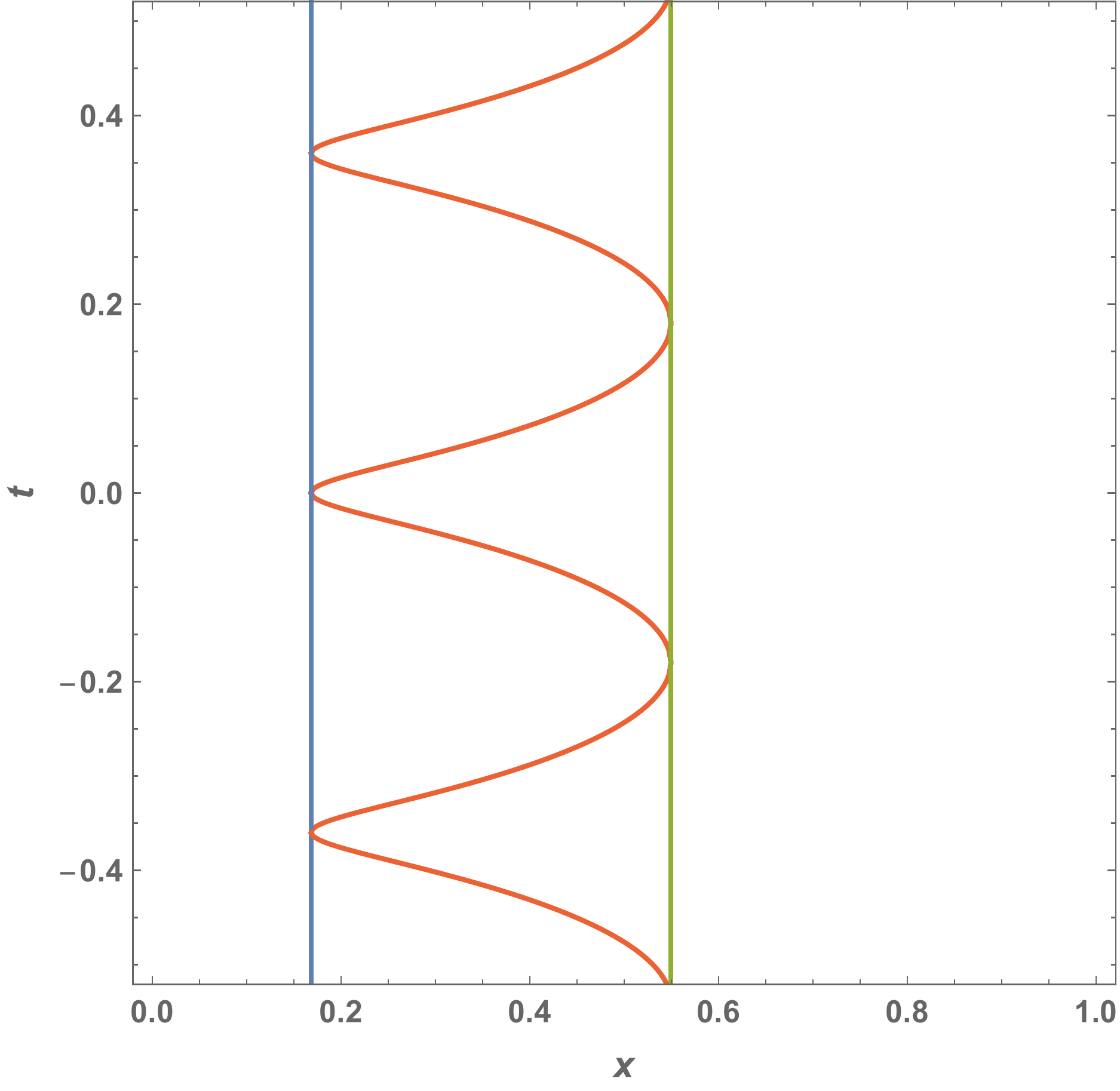}
\includegraphics[width=0.3\textwidth]{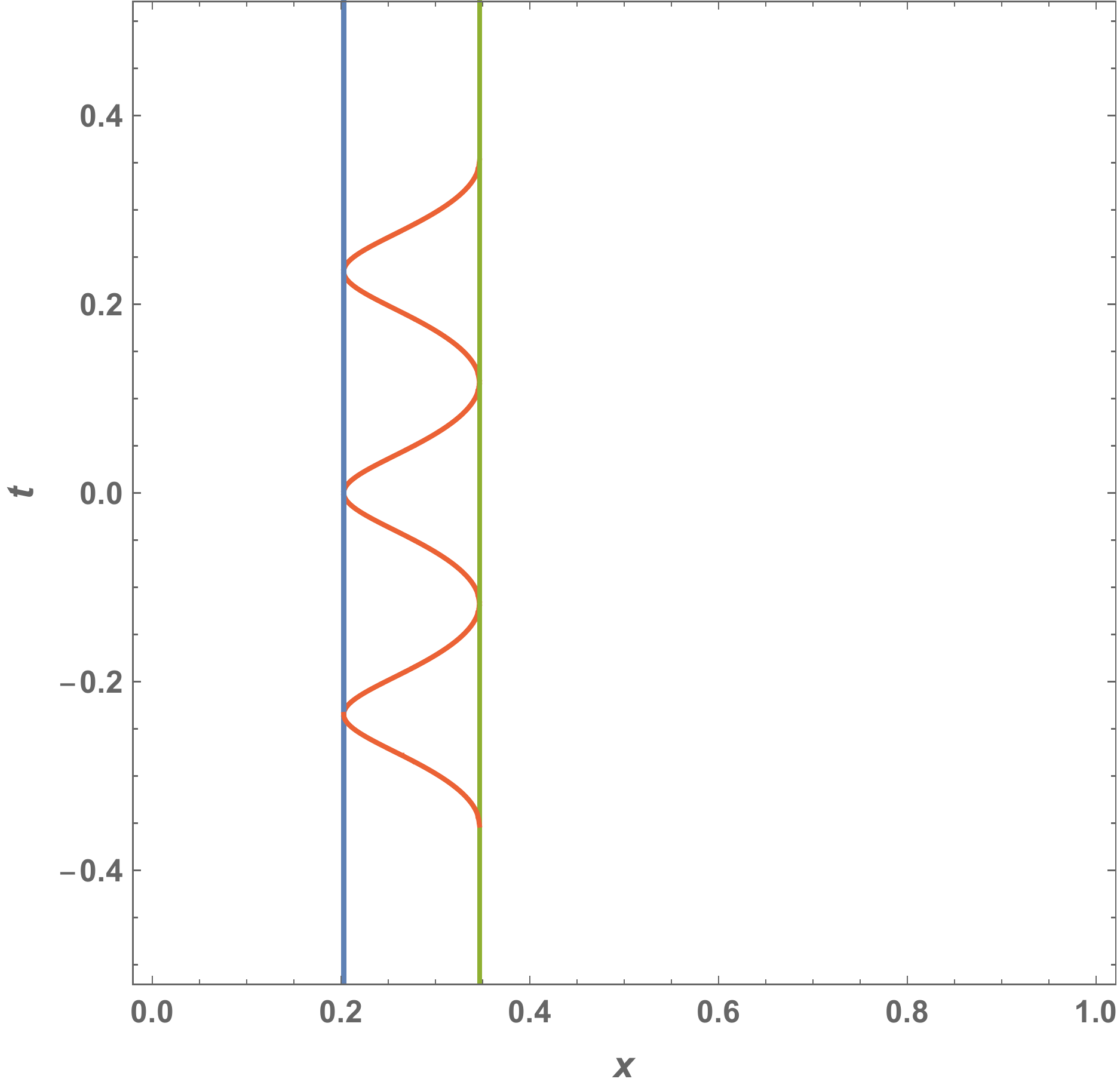}
\caption{\small Plot of the motion $x(t)$ in the curved CK potential with $B=8$, $\ell^2=1$,
and energy values:  $E=-10$ (left), $E=-13$ (center) $E=-16$ (right). The left and right vertical lines are for the values $x_-$ and $x_+$ bounding the motion. The motion in the graphic is limited to three oscillations. \label{fig4}}
\end{figure}

Notice the different sign in the signature of $f_{\epsilon}$. This reflects the fact that all the analytic properties of $f_{\epsilon}$ are reversed from the previous case, so that here we have $a_{1}(x,H)>0$. We will then form the products:

\begin{equation}\label{aepsilon}
    A_{\epsilon}= f_{1}^{\tilde{\gamma}_{\epsilon}(H)}g_{\epsilon},
\end{equation}
where
\begin{equation}
    \tilde\gamma_{\epsilon}(H)= -\frac{2\tilde{\phi}_{1}(H)\sqrt{-H+\epsilon B\sqrt{\kappa}}}{B\sqrt{\kappa}}\, .
\end{equation}
From this and the same computations as in the last section, we deduce that
\begin{equation}\label{braket1}
    \Lambda(A_{\epsilon})= -i \epsilon 
\frac{4\sqrt{\kappa}\sqrt{H^2-B^2\kappa}}{\sqrt{-H+ B\sqrt{\kappa}}-\sqrt{-H- B\sqrt{\kappa}}}
\equiv 
-i\epsilon \alpha(H),
\end{equation}
together with the signature
\begin{equation}
    \Gamma(A_{\epsilon})= -1\,.
\end{equation}
%We have checked from action-angle formulas that 
%\begin{equation}\label{omega}
%\alpha(H)= \omega(H)=\frac{4\sqrt{\kappa}\sqrt{H^2-B^2\kappa}}{\sqrt{-H+B\sqrt{\kappa}}-\sqrt{-H-B\sqrt{\kappa}}}\,,
%\end{equation}
Thus, here we have also obtained fundamental ladder functions (\ref{aepsilon}) with $\omega(H) = \alpha(H)$. The motion (for $B>0$ and $\epsilon=-1$) is given by 
\begin{equation}\label{motion2}
\begin{array}{l}
\displaystyle \tilde\gamma_{-1}(E)\,{\rm arctan}
\left[\frac{p(x,E)}{\frac{\tilde\phi_1(E)}{\tanh x}-\frac{B}{2\tilde\phi_1(E)}}\right]
+
\\[4.ex]
\qquad \qquad \qquad
\displaystyle {\rm arctan} \left[\frac{2p(x,E)\sqrt{-E-B}}{(-B+2\ell^2)\coth x - B- 2(E+\ell^2)} \right]
-\omega(E) t = \theta_0 \, .
\end{array}
\end{equation}
Some examples of motion in the curved KC potential for different values
of the energy are represented in Fig.~\ref{fig4}.

%%%%%%%%%%%%%
%%%%%%%%%%%%%%

\sect{The cases $\ell=0$ and $\kappa \to 0$}

\subsection{The limit $\ell \to 0$}

We will start with the expression (\ref{aepsilon})
and choose the parameter $\epsilon = -1$ which corresponds to the ladder function
$A^-$, according to the sign in (\ref{braket1}).  We will compare
the limit $\ell\to 0$ with that found by the action-angle method \cite{tobepublished}, thus we start with the expression
\begin{equation}\label{aepsilon2}
    A^-\equiv f_{1}^{\tilde{\gamma}_{-1}(H)}g_{-1},
\end{equation}
where
\begin{equation}
g_{-1}= \frac{1}{{\sqrt{\kappa}} (1-\coth {\sqrt{\kappa}}\ r )}
\left( \frac{(- B {\sqrt{\kappa}}+2\ell^2 \kappa)  \coth {\sqrt{\kappa}}\ r - B{\sqrt{\kappa}}-2(H+\ell^2 \kappa)}{2 \sqrt{-H-  B {\sqrt{\kappa}}}} + ip\right)
\end{equation}
and
\begin{equation}
f_1=  {\sinh \sqrt{\kappa}\ r}\left( \frac{\tilde{\phi}_{1}(H)}{\tanh \sqrt{\kappa}\ r}-\frac{B \sqrt{\kappa}}{2 \tilde{\phi}_{1}(H)} + i p\right), \label{definition a KC 2}
\end{equation}
being the exponent 
\begin{equation}
\tilde{\gamma}_{-1}(H)=
  -\frac{2\tilde{\phi}_{1}(H)\sqrt{-H- B\sqrt{\kappa}}}{B\sqrt{\kappa}}\,.
\end{equation}
After some computations and taking $\ell\to 0$, we find
\begin{equation}\label{ladder}
g_{-1}=-\frac{B }{2\,\sqrt{-H-B\sqrt{\kappa}}}\left(\dfrac{1+\coth \sqrt{\kappa}\,r}{1-\coth \sqrt{\kappa}\,r}+\dfrac{2\,H}{B\sqrt{\kappa}\,(1-\coth \sqrt{\kappa}\,r)} +
2\,i\,p\dfrac{\sqrt{-H-B\sqrt{\kappa}}}{B\sqrt{\kappa}\,(1-\coth\sqrt{\kappa} \,r)}\right).
\end{equation}
If we write the other factor in expression (\ref{aepsilon2}) in the exponential form
\begin{equation}
f_{1}^{\tilde{\gamma}_{-1}(H)}=|f_{1}|e^{i \phi(r,p)},
\end{equation}
we  obtain the same expression of the ladder function $A^+$ as in \cite{tobepublished}.

\subsection{The limit $\kappa \to 0$}

If we take the limit $\kappa \to 0$ of the ladder function $A^-$ defined 
in (\ref{aepsilon2}), we should obtain the corresponding 
ladder function of the Kepler-Coulomb system in flat space:
 \begin{equation}
    H_0=p^2 + V_0(r)= p^2-\frac{B  }{ r}+\frac{\ell^2  }{r^2},\qquad 0<r<+\infty \,.
\end{equation}
We can write eq.~(\ref{aepsilon2}) in the following form
\begin{equation}\label{aepsilon3}
    A^+\equiv \left(|f_{1}|\, e^{i\,\arctan\frac{b_1 p}{a_1}}\right)^{\tilde{\gamma}_{-1}(H)}(c_{-1}(r,H) + i d_{-1}(r)\,p)\,.
\end{equation}
After taking the limit $\kappa\to 0$ of this expression, we arrive to the final result
\begin{equation}\label{ladderflat2}
A^{\pm}=F(H_0)\left(-\dfrac{B}{2\sqrt{-H_0}}+\sqrt{-H_0}\,r\mp \,i\,p\,r \right)\,e^{\pm\,i\, \chi(r,p,H_0)}\,,
\end{equation}
where $F(H_0)$ is a function of $H_0$ and
\begin{equation}\label{fflat}
\chi(r,p,H_0)=-\frac{2\sqrt{-H_0}}{B}\,p\,r\,.
\end{equation}
The limit $\kappa \to 0$  the frequency (\ref{omega})  becomes
\begin{equation}\label{wflat}
\omega(H_0)=\dfrac{4}{B}\,(-H_0)^{3/2}\,.
\end{equation}
The factorization $\kappa\to 0$ in terms of these ladder functions becomes
\begin{equation}\label{hfacflat}
-\ell^2=A^+\,A^-+\lambda=r^2\,p^2-B\,r-r^2\,H_0\,,\qquad \lambda=\frac{B^2}{4\,H_0}\,,
\end{equation}
where $H_0=-e$ takes negative values.
They satisfy the Poisson brackets 
\begin{equation}\label{pbaahflat}
\{A^+,A^-\}=\frac{i\,B}{\sqrt{-H_0}}\,,\qquad \{H,A^{\pm}\}=\mp\,i\,\omega(H_0)\,A^{\pm}\,.
\end{equation}

All these expressions are in full agreement with those published in \cite{kuru12}.

\sect{Concluding remarks}

In this paper, we have characterized the ladder functions corresponding to the one dimensional
systems known as Rosen-Morse II and curved Kepler-Coulomb. These two families of
one dimensional systems are the classical version of some of the factorizable solvable systems 
in quantum mechanics \cite{infeld51}. In fact, they were the only ones for which ladder functions were not yet
computed (except the trigonometric counterparts which can be dealt
in the same way, their explicit solutions will be published elsewhere); 
therefore this work constitutes de completion of a program on
the motion and algebraic properties of classical one dimensional systems. One important remark that should be mentioned is that there is a close
connection of ladder functions and action--angle variables
\cite{campoamor12}.

The ladder functions here found are much more complicated than those
of the systems already known. This is reasonable since, for instance, the motion of the KC system in curved space is
much more complicated to describe than that in flat space (which may be found in standard textbooks \cite{goldstein80}). Therefore, we had
to introduce new methods in order to obtain these new ladder functions.
The main results of this paper are summarised in (i) the general formulas 
(\ref{ladderf}) and (\ref{aepsilon}) for the ladder
functions and (ii) the implicit equations for the motion (\ref{motion1}) and (\ref{motion2}) of these systems. We have checked that the formulas here obtained are consistent with
the previous ones known for the P\"oschl-Teller and flat Kepler-Coulomb potentials by
means of appropriate limits.

There is one further comment concerning the freedom of the two possible choices $\epsilon=\pm1$
for the final ladder functions $A_\epsilon$, due to the factor functions $g_\epsilon$.
One way to make the right choice (at least for RMII with $B>0$ and KC with $\kappa\geq 0$), is to pay attention to the complex character of $g_\epsilon(x,H)$:
this function should be complex for $H<-B$, such that if 
$g_{\epsilon} = c_{\epsilon}- id_{\epsilon} p$ as given in (\ref{definition g RM 1}), then its
complex conjugate should be $g_{\epsilon}^* = c_{\epsilon}+ id_{\epsilon} p$  in order to describe bounded motions.
However, the unbounded motions for $H>-B$,
are described by factor functions which satisfy $g_\epsilon(x,H)=  g_\epsilon(x,H)^*$ (up to a global sign) so that they will essentially be real. This is what  happens if $\epsilon=-1$ for both cases RMII and KC above mentioned. This change of character depending on the value of $H$ for the ladder functions was satisfied for all the other simpler cases discussed in \cite{kuru08}. In this respect, it is also known that in quantum mechanics the character of the symmetry algebra depends on the value of the Hamiltonian operator, hence this algebra may change from compact to non compact when the energy values belong to the discrete or to the continuous spectrum.

The one dimensional systems, although not realistic in most physical problems, have considerable interest. For instance, integrable systems can be separated into a set of
one dimensional problems. In particular, many superintegrable systems in higher dimensions
can be separated in some classes of one dimensional systems which have ladder functions.
In this case, the ladder functions together with the so called `shift functions' 
allow to get the symmetries and then, prove the superintegrability in a straightforward way \cite{calzada14,angel16}.

The ladder functions are the classical analog of the ladder operators in quantum 
mechanics. Such ladder operators are used in the construction of quantum coherent states,
while in the classical context the ladder functions allow to characterize the classical motion. Therefore,
the knowledge of such functions and operators constitute a natural approach to
connect classical and quantum systems \cite{campoamor12,kuru07,AHH}. The construction of ladder operators for the quantum RMII and KC systems as well as their coherent states will be investigated in the near future.

 \section*{Acknowledgments}

This work was partially supported by the Spanish MINECO (MTM2014-57129-C2-1-P) and Junta de Castilla y Le\'on and FEDER project  (BU229P18, VA057U16, VA137G18).  \c{S}.~Kuru acknowledges Ankara University and the warm hospitality at Department of Theoretical Physics, University of Valladolid, where part of this work has been done. V. H. acknowledges the support of research grant from NSERC of Canada.

%%%%%%%%%%%%%%%%%%%%%%%%%%

\section*{Appendix: Signature of $f_{\epsilon}$ and $g_{\epsilon}$ for RMII)}

We will determine the signature of the functions $f_{\epsilon}$ and $g_{\epsilon}$ defined by (\ref{definition f RM 1}) and (\ref{definition g RM 1})  with a simple argument by taking into account the fact that $\Gamma(f_{\epsilon})=\frac{a_{\epsilon}(x_+,H)}{a_{\epsilon}(x_-,H)}=\pm 1$ and a similar expression for $\Gamma(g_{\epsilon})$ because of the factorization condition.

The point is to show that for $\epsilon=1$, $a'_{1}(x,H)$ is never equal to zero while for $\epsilon=-1$ $a_{-1}(x,H)$ is never equal to zero, where $'$ denotes the differentiation with respect to $x$. Thus, in the case $\epsilon=1$, the function is strictly increasing or decreasing, which implies it is one to one so that $\frac{a_{\epsilon}(x_+,H)}{a_{\epsilon}(x_-,H)}\neq 1$. In the case $\epsilon=-1$, the function is always positive (or negative), so that $\frac{a_{\epsilon}(x_+,H)}{a_{\epsilon}(x_-,H)}\neq -1$. 

For $\epsilon=1$, we see from the definition (\ref{definition a RN 1}) of $a_{\epsilon}$  that $a'_{1}(x,H)=0$ if and only if $\phi_{1}(H)\cosh x+\frac{B}{2\phi_{1}(H)}\sinh x=0$. This would imply that for a value of $x$ we have $\tanh x= -\frac{2\phi_{1}^2}{B}$. But since $-1<\tanh(x)<1$ this is only satisfied if:
\begin{equation}
    \frac{2\phi_{1}(H)^2}{B}<1. \label{equation satisfied by phi}
\end{equation}
If we develop this equation using eq.~(\ref{definition of Phi RM 1}) for $\phi_{1}$ we see that it cannot be satisfied and thus $a'_{1}(x,H)$ does not cancel and $\Gamma(f_{1})=-1$.

For $\epsilon=-1$, the condition for the cancelation of $a_{-1}(x,H)$ is the same: $\frac{2\phi_{1}(H)^2}{B}<1$. This can be seen from the fact that $a_{-1}(x,H)=a'_{1}(x,H)$, which may be shown by using the relation $\phi_{1}\phi_{-1}=2B$. We thus have $\Gamma(f_{-1})=1$. 
Finaly, we obtain $a_{-1}(x,H)>0$ by evaluating $a_{-1}(x,H)$ at $x=0$.

For the function $g_{\epsilon}$, the argument is simpler since $c'_{\epsilon}(x,H)=\frac{\sqrt{-H+\epsilon B}(\tanh^2 x-1)}{(\tanh x-\epsilon)^2}=0$ implies $\tanh^2 x=1$ or $\epsilon B-H=0$ which is excluded. Thus because $c_{\epsilon}(x,H)$ is one to one,
$\Gamma(g_{\epsilon})=-1$.

\end{document}